\documentclass[onecolumn,twoside]{IEEEtran}
\usepackage{mathpazo}
\usepackage{times}
\usepackage{amsmath}
\usepackage{amsfonts}
\usepackage{latexsym}
\usepackage{amssymb}

\usepackage{upref}
\usepackage{theorem}
\usepackage{graphicx}
\usepackage{psfrag}
\usepackage{cite}

\usepackage{color}

\usepackage[normalem]{ulem}

\theoremstyle{plain}
\theorembodyfont{\normalfont\slshape}

\newtheorem{thm}{Theorem$\!$}
\newenvironment{theorem}
{\begin{thm}\hspace*{-1ex}{\bf.}}{\end{thm}}

\newtheorem{clm}[thm]{Claim$\!$}

\newtheorem{lem}[thm]{Lemma$\!$}
\newenvironment{lemma}{\begin{lem}\hspace*{-1ex}{\bf.}}{\end{lem}}

\newtheorem{prop}[thm]{Proposition$\!$}

\newtheorem{cor}[thm]{Corollary$\!$}
\newenvironment{corollary}{\begin{cor}\hspace*{-1ex}{\bf.}}{\end{cor}}

\newtheorem{defn}[thm]{Definition$\!$}
\newenvironment{definition}{\begin{defn}\hspace*{-1ex}{\bf.}}{\end{defn}}

\newtheorem{xmpl}[thm]{Example$\!$}
\newenvironment{example}{\begin{xmpl}\hspace*{-1ex}{\bf.}}{\hfill $\Box$ \end{xmpl}}

\newtheorem{cnstr}{Construction$\!$}

\newenvironment{construction}{\begin{cnstr}\hspace*{-1ex}{\bf.}}{\hfill$\Box$\end{cnstr}}

\newcounter{enumrom}
\renewcommand{\theenumrom}{(\roman{enumrom})}

\makeatletter
\renewcommand{\@endtheorem}{\endtrivlist}
\makeatother

\makeatletter
\renewcommand{\thefigure}{{\@arabic\c@figure}}
\renewcommand{\fnum@figure}{{\bf Figure\,\thefigure}}
\makeatother

\newcommand{\cB}{\mathcal{B}}

\newcommand{\cD}{\mathcal{D}}

\newcommand{\cF}{\mathcal{F}}

\newcommand{\cL}{\mathcal{L}}

\newcommand{\cP}{\mathcal{P}}

\newcommand{\mathset}[1]{\left\{#1\right\}}
\newcommand{\abs}[1]{\left|#1\right|}

\newcommand{\parenv}[1]{\left( #1 \right)}

\newcommand{\be}[1]{\begin{equation}\label{#1}}
\newcommand{\ee}{\end{equation}}

\renewcommand{\leq}{\leqslant}

\renewcommand{\geq}{\geqslant}

\renewcommand{\Bbb}{\mathbb}

\newcommand{\Cref}[1]{Co\-ro\-lla\-ry\,\ref{#1}}

\renewcommand{\Bbb}{\mathbb}

\newcommand{\N}{{\Bbb N}}
\newcommand{\Q}{{\Bbb Q}}
\newcommand{\R}{{\Bbb R}}
\newcommand{\Z}{{\Bbb Z}}

\newcommand{\cPsi}{\cP_{\mathrm{si}}(\Sigma^k)}

\DeclareMathOperator{\fr}{fr}
\DeclareMathOperator{\sub}{sub}
\DeclareMathOperator{\scl}{cl}
\DeclareMathOperator{\sint}{int}

\DeclareMathOperator{\im}{Im}
\DeclareMathOperator{\ess}{ess}
\DeclareMathOperator{\suffchop}{SuffChop}

\newcommand{\ccap}{\mathsf{cap}}

\newcommand{\limup}[1]{\lim_{#1\rightarrow\infty}}

\newcommand{\emptyword}{\varepsilon}
\newcommand{\esubseteq}{\subseteq^e}

\outer\def\proclaim #1. #2\par{\medbreak
 \noindent{\bf#1.\enspace}{\sl#2\par}%
 \ifdim\lastskip<\medskipamount \removelastskip\penalty55\medskip\fi}

\begin{document}

\title{\textbf{Encoding Semiconstrained Systems}}

\author{\large
Ohad Elishco,~\IEEEmembership{Student Member,~IEEE},
Tom Meyerovitch,
Moshe~Schwartz,~\IEEEmembership{Senior Member,~IEEE}
\thanks{
  The material in this paper was presented in part at the
  IEEE International Symposium on Information
  Theory (ISIT 2016), Barcelona, Spain, July 2016.}%
\thanks{Ohad Elishco is with the Department
  of Electrical and Computer Engineering, Ben-Gurion University of the Negev, Beer Sheva 8410501, Israel
   (e-mail: ohadeli@post.bgu.ac.il).}%
\thanks{Tom Meyerovitch is with the Department of Mathematics, Ben-Gurion University of the Negev,
   Beer Sheva 8410501, Israel (e-mail: mtom@math.bgu.ac.il).}%
\thanks{Moshe Schwartz is with the Department
   of Electrical and Computer Engineering, Ben-Gurion University of the Negev,
   Beer Sheva 8410501, Israel
   (e-mail: schwartz@ee.bgu.ac.il).}%
\thanks{This work was supported in part by the People Programme (Marie Curie Actions) of the European Union's Seventh Framework Programme (FP7/2007-2013) under REA grant agreement no.~333598
and by the Israel Science Foundation (grant no.~626/14).} 
}

\maketitle

\begin{abstract}
  Semiconstrained systems were recently suggested as a generalization
  of constrained systems, commonly used in communication and
  data-storage applications that require certain offending
  subsequences be avoided. In an attempt to apply techniques from
  constrained systems, we study sequences of constrained systems that
  are contained in, or contain, a given semiconstrained system, while
  approaching its capacity. In the case of contained systems we
  describe two such sequences resulting in constant-to-constant
  bit-rate block encoders and sliding-block encoders. Perhaps surprisingly, in
  the case of containing systems we show, under commonly-made
  assumptions, that the only constrained system that contains a given
  semiconstrained system is the entire space. A refinement to this
  result is also provided, in which semiconstraints and zero
  constraints are mixed together.
\end{abstract}


\section{Introduction}

\IEEEPARstart{M}{any} communication and data-storage systems employ
constrained coding. In such a scheme, information is encoded in
sequences that avoid the occurrence of certain subsequences. Perhaps
the most common example is that of $(d,k)$-RLL which comprises of
binary sequences that avoid subsequences of $k+1$ $0$'s, as well as
two $1$'s that are separated by less than $d$ $0$'s. For various other
examples the reader is referred to \cite{Imm04} and the many
references therein.

The reason for avoiding such subsequences is mainly due to the fact
that their appearance contributes to noise in the system. However, by
altogether forbidding their occurrence, the possible rate at which
information may be transmitted is severely reduced. By relaxing the
constraints and allowing some appearances of the offending
subsequences, the rate penalty may be reduced. So rather than imposing
combinatorial constraints on all substrings of the output, we impose
statistical constraints on substrings that are sampled from the output
at a uniform random offset. Such an approach was studied, for example,
in the case of channels with cost constraints
\cite{KarNeuKha88,KhaNeu96}.

A general approach was suggested in \cite{EliMeySch16}, in which a
\emph{semiconstrained system (SCS)} was defined by a list of offending
subsequences, and an upper bound (called a \emph{semiconstraint}) on
the frequency of each subsequence appearing. Note that constrained
systems (which we call \emph{fully constrained} for emphasis) is a
special case of semiconstrained systems, in which only semiconstraints
of frequency $0$ are used.

A careful choice of semiconstraints also allows the study of systems
that, up to now, were studied in an ad-hoc manner only. As examples we
mention DC-free RLL coding \cite{Kur11}, constant-weight ICI coding
for flash memories \cite{KaySie14,QinYaaSie14,CheChrKiaLinNguVu16a,CheChrKiaLinNguVu16b}, and coding to mitigate
the appearance of ghost pulses in optical communication
\cite{ShaSkiTur10,ShaTurTur07}.

One of the most important questions, given a SCS, is how to encode any
unconstrained input sequence into a sequence that satisfies all the
given semiconstraints. The various encoding schemes suggested in
\cite{Kur11,KaySie14,QinYaaSie14,ShaSkiTur10,ShaTurTur07,CheChrKiaLinNguVu16a}
are ad-hoc and do not apply to general SCS. The encoding scheme for
channels with cost constraints given in \cite{KhaNeu96} (which overlap
somewhat with SCS) is indeed general, however it is not capacity
achieving. Later, within the scope of channels with cost constraints,
and motivated by partial-response channels,
\cite{SorSie06,KraKarYanWil14} briefly report on capacity-achieving
schemes, however, not in the full generality we consider in this
paper.

Under the assumption that the input stream consists of
i.i.d.~uniformly-random bits, a general capacity-achieving encoding
scheme for SCS was described in \cite{EliMeySch16}. The
scheme involved a maxentropic Markov chain over a modified De-Bruijn
graph. Input symbols were converted via an arithmetic decoder to a
biased stream of symbols which were used to generate a path in the
graph, which in turn generated symbols to be transmitted. A reverse
operation was employed at the receiving side. Additionally to the
assumption on the distribution of the input, to enforce a
constant-to-constant bit rate, the encoder has a probability of
failure (albeit, asymptotically vanishing). Thus, not all input
streams may be converted to semiconstrained sequences.

Compared with SCS, for ``conventional'' fully constrained systems
there is a general method for constructing encoders working
arbitrarily close to capacity: The celebrated state-splitting
algorithm. However, as we explain in the following sections, this
method fails even on very simple SCS, due to the fact that in most
cases they do not form regular languages.

In this work we consider the problem of encoding an arbitrary input
string into a sequence that satisfies all the given
semiconstraints. We do not make statistical or combinatorial
assumptions on the input, only that it is sufficiently
long. Specifically, we show the following: For every given SCS that
satisfies certain mild assumptions and every $\epsilon >0$ we present
a fully constrained system that is ``eventually-contained'' in the
given SCS, with a capacity penalty of at most $\epsilon$. This allows
us to construct either block encoders or sliding-block encoders,
trading encoder anticipation for number of states. In the other
direction, we show that no fully constrained system can contain a
given SCS (under certain mild assumptions).  We also observe that any
encoding scheme for a SCS that works for arbitrary input and has both
finite memory and finite anticipation must produce sequences that
satisfy some fully constrained system.

The paper is organized as follows. In Section \ref{sec:defs} we
present the definition and notation used throughout the paper. In
Section \ref{sec:below} we study sequences of constrained systems that
are contained in a given SCS and approach its capacity from below. In
Section \ref{sec:above} we do the reverse, and study constrained
systems containing a given SCS. In the first sections of this paper we make some  assumptions on the SCS under discussion, in particular that they are \emph{fat} (see Definition \ref{def:fat}). These assumptions in particular exclude (classical) fully-constraint systems. In Section \ref{sec:RF} we study more general SCS,  that in particular also allow  fully-constraint systems. We
present conclusions and further research in Section \ref{sec:conc}.

\section{Preliminaries}
\label{sec:defs}

\subsection{Semiconstrained Systems}

Let $\Sigma$ be a finite alphabet and let $\Sigma^*$ denote the set of
all the finite sequences over $\Sigma$. The elements of $\Sigma^*$ are
called \emph{words} (or \emph{strings}). The \emph{length} of a word
$\omega\in\Sigma^*$ is denoted by $\abs{\omega}$. Given two words,
$\omega,\omega'\in\Sigma^*$, their concatenation is denoted by
$\omega\omega'$. Repeated concatenation is denoted using a
superscript, i.e., for any natural $m\in\N$, $\omega^m$ denotes
$\omega^m=\omega\omega\dots\omega$, where $m$ copies of $\omega$ are
concatenated. By convention, $\omega^0=\emptyword$, where $\emptyword$
the unique empty word of length $0$. By extension, if
$S\subseteq\Sigma^*$ is a set of words, then $S^m$ denotes the set
\[ S^m = \mathset{ \omega_1 \omega_2 \dots \omega_m : \forall i, \omega_i\in S},\]
with $S^0=\mathset{\emptyword}$, $S^* = \bigcup_{i\geq 0} S^i$,
and $S^+ = \bigcup_{i\geq 1} S^i$.

The set of  $k$-length subwords of $\omega$ is
defined by
\[ \sub_k(\omega) = \mathset{ \beta\in\Sigma^k ~:~  \omega=\alpha\beta\gamma \text{ for some } \alpha,\gamma\in\Sigma^*}.\]
For $\omega \in \Sigma^*$ and $k \leq \abs{\omega}$, $\fr_\omega^k \in
\cP(\Sigma^k)$ is defined as the uniform measure on $\sub_k(\omega)$
(taken in the multiset sense), where $\cP(\Sigma^k)$ denotes the set
of all probability measures on $\Sigma^k$. We can naturally identify
\[ \cP(\Sigma^k) =  \mathset{ \eta\in [0,1]^{\Sigma^k} : \sum_{\phi\in\Sigma^k}\eta(\phi)=1}.\]
It follows that for all $\beta\in\Sigma^k$,
\[\fr_\omega^k(\beta) = \frac{1}{\abs{\omega}-k+1}\abs{\mathset{(\alpha,\gamma)
    : \alpha,\gamma\in\Sigma^*, \alpha\beta\gamma = \omega}}.\]

\begin{definition}
  Let $\Gamma\subseteq\cP(\Sigma^{k})$ be a set of probability
  measures. We say $\Gamma$ is a \emph{semiconstrained system (SCS)},
  and we define the set of \emph{admissible words for $\Gamma$} by
\[\cB{(\Gamma)}=\mathset{ \omega\in\Sigma^* :  \fr_{\omega}^k\in \Gamma }.\]
\end{definition}

For convenience we also define the set of admissible words of length
exactly $n$ as
\[\cB_n(\Gamma) = \cB{(\Gamma)}\cap \Sigma^n.\]

An important figure of merit we associate with any set of words
$S\subseteq \Sigma^*$ is its capacity.
\begin{definition}
  Let $\Sigma$ by a finite alphabet and $S\subseteq \Sigma^*$. The
  \emph{capacity} of $S$, denoted $\ccap(S)$, is defined as
  \[
  \ccap(S)=\limsup_{n\rightarrow \infty} \frac{1}{n}\log_2 \abs{S\cap\Sigma^n}.
  \]
\end{definition}

Thus, in the case of a SCS $\Gamma$, the capacity $\ccap(\cB(\Gamma))$
intuitively measures the exponential growth rate of the number of
words that satisfy the constraints given by $\Gamma$ as a function of
the word length.

A relaxation of semiconstrained systems was also suggested in
\cite{EliMeySch16}.

\begin{definition}
Let $\Gamma\subseteq\cP(\Sigma^k)$ be a set of probability
measures. For $\epsilon>0$ we denote by $\Gamma^\epsilon$ the set
\[\Gamma^\epsilon = \mathset{\eta\in\cP(\Sigma^k) ~:~
  \inf_{\mu\in\Gamma}\|\eta-\mu\|_{\infty} \leq \epsilon}\]
where
$\|\cdot\|_{\infty}$ is the $\ell_\infty$-norm. The set of
\emph{weakly-admissible words for $\Gamma$} is defined by
\[\overline{\cB}{(\Gamma)}=\mathset{ \omega\in\Sigma^* :  \fr_\omega^k\in \Gamma^{\xi(\abs{\omega})} },\]
where $\xi:\N\to\R^+$ is a function satisfying both $\xi(n)=o(1)$ and
$\xi(n)=\Omega(1/n)$. Also $\overline{\cB}_n(\Gamma)=\overline{\cB}{(\Gamma)}\cap \Sigma^n$.
\end{definition}

We note that $\overline{\cB}(\Gamma)$ was called a \emph{weak
  semiconstrained system (WSCS)} in \cite{EliMeySch16} and was defined
in a slightly different manner, though we shall prefer to use the term
weakly-admissible words for $\Gamma$. It was also shown there that
though it is possible to constrain words of different lengths, it
suffices to consider only words in $\Sigma^k$, i.e., all the offending
patterns are of the same length $k$. Here, since
$\Gamma\subseteq\cP(\Sigma^k)$, the assumption that all the offending
patterns are of length $k$ is implied.

A particular set of probability measures of interest to us is the set
of \emph{shift-invariant probability measures}. We say
$\eta\in\cP(\Sigma^k)$ is \emph{shift-invariant} if for all
$\phi\in\Sigma^{k-1}$,
\[ \sum_{a\in\Sigma}\eta(a\phi)=\sum_{a\in\Sigma}\eta(\phi a).\]
We denote the set of shift-invariant probability measures by
$\cPsi$, which is a closed subset of
$\cP(\Sigma^k)$. These are precisely the probability measures that
arise as marginals of shift-invariant measures on
$\Sigma^{\N}$ or $\Sigma^{\Z}$. For a discussion see
\cite{Hoc2012}. In particular, we have the following
lemma.

\begin{lemma}
  \label{lem:onlysi}
  Fix a finite alphabet $\Sigma$, and $k\geq 2$. If $\Gamma\subseteq
  \cP(\Sigma^k)\setminus \cPsi$ is closed then
  $\ccap(\cB(\Gamma))=-\infty$, i.e., $\cB(\Gamma)$ is a finite set.
\end{lemma}
\begin{IEEEproof}
  For any $\omega\in\Sigma^*$, $\abs{\omega}\geq k$, and any
  $\phi\in\Sigma^{k-1}$, by simple counting
  \[ \abs{\sum_{a\in\Sigma}\fr_\omega^{k}(a\phi)-\sum_{a\in\Sigma}\fr_\omega^{k}(\phi a)}\leq \frac{1}{\abs{\omega}}.\]
  Thus, $\fr_{\omega}^k$ gets arbitrarily close to a shift-invariant
  probability measure as $\abs{\omega}\to\infty$. Since $\cPsi$ and
  $\Gamma$ are closed, there is a positive distance between the
  sets. Therefore, there exists $n\in\N$ such that for all
  $\omega\in\Sigma^*$, $\abs{\omega}\geq n$, we have $\omega\not\in
  \cB(\Gamma)$, i.e., $\cB(\Gamma)$ is finite.
\end{IEEEproof}

Lemma \ref{lem:onlysi} motivates us to study probability measures that
are shift invariant. Another crucial property of a set of probability
measures is given in the following definition.

\begin{definition}\label{def:fat}
  For a set $\Gamma' \subseteq \cPsi$ we denote by $\sint(\Gamma')$
  the interior of $\Gamma'$, and by $\scl(\Gamma')$ the closure of
  $\Gamma'$, both relatively to $\cPsi$. We say $\Gamma \subseteq
  \cP(\Sigma^k)$ is \emph{fat} if
  \[\scl(\sint(\Gamma \cap \cPsi))=\scl(\Gamma \cap \cPsi).\]
\end{definition}


We recall the  following result from \cite{EliMeySch16}:

\begin{theorem}
  \label{th:prevpaper}
  Let $\Gamma\subseteq\cP(\Sigma^k)$ be closed and convex. If $\Gamma$
  is fat then
  \[
  \ccap(\cB(\Gamma))=\ccap(\overline{\cB}(\Gamma))
  =\log_2 \abs{\Sigma}-\inf_{\eta\in\Gamma\cap\cPsi}H(\eta|\eta'),
    \]
  where $H(\cdot|\cdot)$ is the relative entropy function, and
  $\eta'(\phi a) = \sum_{a'\in \Sigma} \eta(\phi a')/\abs{\Sigma}$,
  for all $\phi\in\Sigma^{k-1}$ and $a\in\Sigma$. Additionally,
  $\ccap(\cB(\Gamma))$ and $\ccap(\overline{\cB}(\Gamma))$ are continuous and convex in $P$, and the limits in their definitions exist.
\end{theorem}

\subsection{Fully Constrained Systems}

As noted in the introduction, ``conventional'' constrained
systems are a special case of semiconstrained systems. They can be
viewed as SCS $\Gamma$, where $\Gamma$ is of the form
\[ \Gamma=\mathset{\eta\in\cP(\Sigma^k) : \forall \phi\in\Sigma^k, \eta(\phi) \leq c_\phi},\]
where $c_\phi\in \mathset{0,1}$ for all $\phi\in\Sigma^k$. In other
words, every substring of length $k$ is either completely forbidden,
or unconstrained. We will refer to those as \emph{fully constrained
  systems} and denote a set $\Gamma$ of this form as
$\Gamma_{\mathset{0,1}}$.

Let $G=(V,E)$ be a finite directed graph, where we allow parallel
edges. A labeling function $\cL:E\to\Sigma^q$ assigns a length-$q$
label over the alphabet to each edge. By simple extension, the label
of a directed (non-empty) path in the graph $\gamma=e_1 \to e_2
\to\dots \to e_n$ is defined as
$\cL(\gamma)=\cL(e_1)\cL(e_2)\dots\cL(e_n)$. Finally, we define the
language presented by the graph $G$, denoted $\cL(G)$, to be the
labels of all finite directed paths in $G$.

Constrained systems have been widely studied \cite{LinMar85,Imm04}. In
particular, it is well known that in case $\Gamma$ is of the form $\Gamma_{\mathset{0,1}}$, $\cB(\Gamma)=\cL(G)$ for some finite directed labeled graph $G$ in the manner described above. An immediate consequence is the fact that $\cB(\Gamma)$ is a regular language in the Chomsky hierarchy of formal languages \cite{Sha08}. We do note, however, that not all regular languages (which correspond to languages of sofic subshifts) are constrained systems (which are defined by a finite number of forbidden words, and correspond to subshifts of finite type).

A wide variety of tools exist for manipulating constrained systems,
including the state-splitting algorithm (see \cite[Chapter $5$]{LinMar85}). In
essence, under mild assumptions, given a constrained system
$\cB(\Gamma)=\cL(G)$, and two positive integers $p$ and $q$ that
satisfy $p/q < \ccap(\cB(\Gamma))$, we can find another constrained
system $\cB'(\Gamma)=\cL(G')$, an \emph{encoder}, with the following
properties:
\begin{itemize}
\item
  $\cL(G')\subseteq \cL(G)$.
\item
  $\ccap(\cB'(\Gamma))=p/q$, also called the \emph{rate} of the encoder.
\item
  $G'$ is a $p:q$ encoder for $\cL(G)$ with finite anticipation
  $a\in\N\cup\mathset{0}$, i.e., the out-degree of each vertex is
  $2^p$, the edges labels in $G'$ are from $\Sigma^q$, and paths of
  length $a+1$ that start from the same vertex and generate the same
  word agree on the first edge.
\end{itemize}

Unfortunately, even for very simple semiconstraints, $\cB(\Gamma)$ is
not a regular language in general. As an example, for
$\Sigma=\mathset{0,1}$, and $\Gamma$ such that for all $\mu\in\Gamma$,
$\mu(1)\leq p$, it is easily seen that for any rational $0<p<1$, the
semiconstrained system $\cB(\Gamma)$ is a non-regular context-free
language, whereas for any irrational $p$ the system is not even
context free \cite[\S 4.9, Exercise 25]{Sha08}. Thus, the wonderful
machinery of the state-splitting algorithm cannot be applied directly
for general SCS.

Another important property of languages associated with
fully constrained systems is that these languages are
\emph{factorial}. This means that a subword of an admissible word is
also an admissible word. Factoriality implies for instance that if
$\Gamma$ is with the property $\Gamma_{\mathset{0,1}}$, the sequence
$\frac{1}{n}\log\abs{\cB(\Gamma_{\mathset{0,1}})}$ is subadditive, so
the $\limsup$ in the definition of the capacity is actually a limit by
Fekete's Lemma.  The factoriality property is not shared by SCS in
general.

\section{Approaching Capacity from Below}
\label{sec:below}

In this section we study the problem of finding a sequence of
fully constrained systems that are contained in a given
semiconstrained (or weakly semiconstrained) system, with the
additional requirement that the capacity of the former approaches that
of latter in the limit. We present two such sequences which induce
(perhaps after state splitting) two possible encoders for the SCS or
WSCS.

Before continuing on, we pause to consider what properties we require
of an encoder. An encoder is nothing more than a function
$\phi:\Sigma^{\N}\to X$ for translating an unconstrained sequence of
input symbols $\Sigma^\N$, into another sequence obeying a given set
of constraints, $X\subseteq\Sigma^{\N}$. A general encoder for SCS was
already described in \cite{EliMeySch16}. However, that encoder had a
probability of failure, i.e., it would not work on some input
sequences. We are therefore interested in finding an encoder that
always succeeds.

Furthermore, we would like this encoder to have finite memory and
anticipation, intuitively described as having every encoded output
symbol depend on finitely many input symbols preceding it (memory),
and finitely many input symbols following it (anticipation). Once we
require finite memory and anticipation, an encoder unavoidably becomes
one for a fully constrained system.  The following lemma shows that
any encoding function with finite memory and finite anticipation can
be presented by a graph.  A version of the following lemma can be
found in \cite[Chapter 3]{LinMar85}, and we bring here a short proof
for completeness.
\begin{lemma}
  \label{regularlang}
  Let $X\subseteq\Sigma^{\N}$ be such that it has an encoder
  $\phi:\Sigma^{\N}\to X$ with finite anticipation and finite memory.
  Then $\im(\phi)$ can be presented by a graph.
\end{lemma}

\begin{IEEEproof}
  Let $x,y\in\Sigma^{\N}$ and define the metric $d(x,y)=2^{-n(x,y)}$,
  where $n(x,y)$ is the first coordinate $j$ for which $x_j\neq y_j$.
  Note that if $\phi$ has a finite memory and finite anticipation it
  is continuous with respect to $d(\cdot,\cdot)$.  The metric
  $d(\cdot,\cdot)$ on $\Sigma^{\N}$ generates the product topology
  under which the space $\Sigma^{\N}$ is compact.  the image of a
  compact set under a continuous function is compact and therefore
  $\im(\phi)$ is compact.  Moreover, $\Sigma^{\N}$ is a shift of
  finite type, thus $\im(\phi)$ is a sofic shift, i.e., it can be
  presented by a graph.
\end{IEEEproof}

Thus, in what follows, we focus on studying fully constrained systems
contained in a given SCS. One of our goals is to determine the
following function.

\begin{definition}
  \label{def:below}
  Let $\Gamma\subseteq\cP(\Sigma^k)$ be a SCS.
  The \emph{fully constrained contained capacity} is defined as
  \[ \ccap^\subseteq(\Gamma) = \sup_{\cL(G)\subseteq\cB(\Gamma)} \ccap(\cL(G)).\]
\end{definition}

It will be easier for us to describe fully constrained systems that
are only \emph{eventually} contained in the desired SCS. Formally,
given two infinite subsets, $S_1,S_2\in\Sigma^*$, we say $S_1$ is
eventually contained in $S_2$, denoted $S_1\esubseteq S_2$, if
$\abs{S_1 \setminus S_2} < \infty$.  A fully constrained system that
is eventually contained in a given SCS may easily be transformed into
another fully constrained system that is contained (in the usual
sense) in the given SCS by removing the words that are inadmissible in
the SCS.

\subsection{Block Encoders for SCS}

The first sequence of fully constrained systems we construct are each
presented by a graph with a single state. Such graphs are called block
encoders.

Let $\Gamma$ be a fat SCS. The fat condition on
$\Gamma$ guarantees that it can be slightly shrunk while
remaining not empty. More formally, for any $\epsilon>0$
we define the set $\Gamma_{\epsilon}$ by
\begin{equation}
  \label{eq:gammaeps}
  \Gamma_{\epsilon}= \mathset{\eta\in\cP(\Sigma^k) ~:~ \inf_{\mu\in\Gamma^c} \|\eta-\mu\|_{\infty}>\epsilon}
\end{equation}
where $\|\cdot\|_{\infty}$ is the $\ell_\infty$-norm.
	
If $\Gamma$ is fat then there exists $\epsilon>0$ such that
$\Gamma_{\epsilon}\neq\emptyset$ and $\Gamma_{\epsilon}$ is also
fat. It is also obvious that $\Gamma_{\epsilon} \subseteq \Gamma$. We
say such an $\epsilon$ is \emph{$\Gamma$-admissible}.

Note that in the definition of $\Gamma_{\epsilon}$ we consider
$\mu\in\cP(\Sigma^k)$ as a vector of numbers and use the
$\ell_\infty$-norm instead of the usual total-variation norm. 
The particular choice of norm is a side issue and does not significantly change the essential results.   
\begin{construction}
  \label{con:A}
Let $\Gamma$ be a SCS. For every $m\in\N$ we construct $R_m(\Gamma)\subseteq\Sigma^*$ by defining
\[ R_m(\Gamma) = \cB_m(\Gamma)^*.\]
\end{construction}

By definition, $R_m(\Gamma)$ from Construction \ref{con:A} is a
regular language. It may be presented as the language of the following
graph $G$: the graph contains a single vertex, all the edges are self
loops and are labeled by the words of $\cB_m(\Gamma)$, i.e., the
length-$m$ words in $\cB(\Gamma)$.

\begin{theorem}
\label{th:rep1}
Let $\Gamma$ be a convex fat SCS. Then for any $\Gamma$-admissible
$\epsilon>0$, there exists $M_\epsilon\in\R$ such that for all
$m>M_\epsilon$
\[ R_m(\Gamma_{\epsilon})\subseteq \cB(\Gamma).\]
\end{theorem}

\begin{IEEEproof}
First we mention that if $\Gamma$ is convex then so is
$\Gamma_{\epsilon}$.  Consider $\omega\in R_m(\Gamma_{\epsilon})$ and
write $\omega=\omega_1 \omega_2\dots \omega_{\ell}$ with
$\omega_i\in\cB_m(\Gamma_{\epsilon})$.  For any $\phi\in\Sigma^k$, we
bound the number of occurrences of $\phi$ in $\omega$.

For every $1\leq i\leq \ell$, we denote
\[ \mu_i=\fr_{\omega_i}^k\in\Gamma_{\epsilon}.\]
Every $\phi\in\Sigma^k$ appears in $\omega_i$ exactly
$\mu_i(\phi)(m-k+1)$ times. Additionally, in each concatenation point
between $\omega_i$ and $\omega_{i+1}$, the word $\phi$ can appear at
most another $k-1$ times.

Since $\Gamma$ is convex, $\Gamma_{\epsilon}$ is also convex, and then
\[ \mu=\frac{1}{\ell}\sum_{i=1}^{\ell}\mu_i \in \Gamma_{\epsilon}.\]
It now follows that
\begin{align}
	\label{eq:rhsblock1}
	\fr_\omega^k(\phi) &\leq
	\frac{\sum_{i=1}^{\ell} \mu_i(\phi)(m-k+1)+ (\ell-1)(k-1)}{m\ell-k+1} \nonumber \\
	&=\frac{\mu(\phi)\ell(m-k+1)+(\ell-1)(k-1)}{m\ell-k+1}\nonumber \\
	&=\mu(\phi) +\frac{(\ell-1)(k-1)(1-\mu(\phi))}{m\ell-k+1} \nonumber\\
	&\leq \mu(\phi) +\frac{(\ell-1)(k-1)}{m\ell-k+1}.
\end{align}
Additionally,
\begin{align}
	\fr_\omega^k(\phi) &\geq
	\frac{\sum_{i=1}^{\ell}\mu_i(\phi)(m-k+1)}{m\ell-k+1}\nonumber \\
	&=\frac{\mu(\phi)\ell(m-k+1)}{m\ell-k+1}\nonumber \\
	&=\mu(\phi)
	-\frac{(\ell-1)(k-1)}{m\ell-k+1}.\label{eq:rhsblock2} 
\end{align}

Following the right-hand side of \eqref{eq:rhsblock1} and of \eqref{eq:rhsblock2} we can continue the analysis assuming all $\omega_i$ share the same measure $\mu\in\Gamma_{\epsilon}$. 
Now define
\[L(m,\ell)=\frac{(\ell-1)(k-1)}{m\ell-k+1},\]
and thus,
\[ \abs{\fr_\omega^k(\phi)-\mu(\phi)} \leq L(m,\ell).\]
We note that for every $m>k-1$, $\ell\geq 2$, the function $L(m,\ell)$
is monotone increasing in $\ell$. It follows that
\[ L(m,\ell) < \lim_{\ell\to\infty} L(m,\ell) = \frac{k-1}{m}.\]
Hence, we can take
\begin{equation}
  \label{eq:MeA}
  M_{\epsilon}=\frac{k-1}{\epsilon},
\end{equation}
and obtain that for every $m>M_{\epsilon}$,
\[ L(m,\ell) < \epsilon.\]
The calculation holds for every $\phi\in\Sigma^k$. Thus, we showed
that for every $m>M_{\epsilon}$, every $\phi\in\Sigma^k$, and every
$\omega\in R_m(\Gamma_{\epsilon})$, we have $\fr_{\omega}^k \in
\Gamma$, and therefore $R_m(\Gamma_{\epsilon})\subseteq \cB(\Gamma)$.
\end{IEEEproof}

We observe that $M_{\epsilon}=\Omega(\frac{1}{\epsilon})$.
The following theorem shows that the sequence of systems
$R_m(\Gamma_{\epsilon})$ has a capacity that approaches
$\ccap(\cB(\Gamma_{\epsilon}))$ as $m$ grows.

\begin{theorem}
  \label{th:rep2}
  Let $\Gamma$ be a closed convex fat SCS. Then for every
  $\Gamma$-admissible $\epsilon> 0$ the following limit exists and
  \[\limup{m} \ccap(R_m(\Gamma_{\epsilon}))=\ccap(\cB(\Gamma_{\epsilon})).\]
\end{theorem}

\begin{IEEEproof}
  We observe that
  \[ \abs{R_m(\Gamma_{\epsilon})\cap\Sigma^n}=
  \begin{cases}
    \abs{\cB_m(\Gamma_{\epsilon})}^{\frac{n}{m}} & \text{if $m|n$,}\\
    0 & \text{otherwise.}
  \end{cases}
  \]
  It follows that
  \begin{align*}
    \ccap(R_m(\Gamma_{\epsilon})) &= \limsup_{n\to\infty} \frac{1}{n}\log_2\abs{R_m(\Gamma_{\epsilon})\cap\Sigma^n} \\
    &= \frac{1}{m}\log_2 \abs{\cB_m(\Gamma_{\epsilon})}.
  \end{align*}
  Hence,
  \begin{align*}
    \limsup_{m\to\infty}\ccap(R_m(\Gamma_{\epsilon})) & =\limsup_{m\to\infty}\frac{1}{m}\log_2 \abs{\cB_m(\Gamma_{\epsilon})} \\
    &= \ccap(\cB(\Gamma_{\epsilon})),
  \end{align*}
  by the definition of capacity. However, since $\epsilon$ is
  $\Gamma$-admissible, we have a fat $\Gamma_{\epsilon}$. By
  Theorem \ref{th:prevpaper}, the limit in the definition of the
  capacity for the SCS exists, which completes the proof.
\end{IEEEproof}

We note that if $\epsilon_1\leq\epsilon_2$ and $\epsilon_2$ is $\Gamma$-admissible, then $\epsilon_1$ is also
$\Gamma$-admissible. 

\begin{corollary}
  \label{cor:belowblock}
  For any SCS with a closed convex fat $\Gamma$ there exist block
  encoders with rate arbitrarily close to $\ccap(\cB(\Gamma))$.
\end{corollary}

\begin{IEEEproof}
  By Theorem \ref{th:prevpaper} the limit in the capacity definition
  exists and the capacity, which is given by the relative entropy
  function, is continuous with respect to the restrictions. Thus,
  \[\lim_{\epsilon\to 0}\ccap(\cB(\Gamma_{\epsilon})) = \ccap(\cB(\Gamma)),\]
  where $\epsilon$ is $\Gamma$-admissible. It follows that Theorem
  \ref{th:rep1} and Theorem \ref{th:rep2} show that it is possible to
  build a block encoder to a given SCS with rate arbitrarily close to
  $\ccap(\cB(\Gamma))$.
\end{IEEEproof}

While the block encoders we constructed are quite simple, and have
rate $p/q$ arbitrarily close to $\ccap(\cB(\Gamma))$, we do however point a major drawback. The edges are labeled by words from
$\Sigma^m$. Thus, the encoder is not $p:q$ but $mp:mq$. For a fair
comparison with the next construction, if we transform this to an
encoder with labels from $\Sigma$ (e.g., via a standard tree
argument), the anticipation becomes $\Omega(m)$, which is undesirable.

\subsection{Sliding-Block Encoders}

Unlike Construction \ref{con:A}, in which a sequence was a concatenation of independent blocks, the construction we now present has a sliding-window restriction.

\begin{construction}
  \label{con:B}
  Let $\Gamma$ be a SCS. For every $m\in\N$ we construct $N_m(\Gamma)\subseteq\Sigma^*$ by defining
  \[N_m(\Gamma) = \mathset{ \omega\in\Sigma^* : 
  	\sub_m(\omega)\subseteq \cB(\Gamma)}.\]
\end{construction}

We observe that $N_m(\Gamma)$ from Construction \ref{con:B} is a
fully constrained system. Indeed, it is defined by a finite set of
forbidden words, $\Sigma^m\setminus \cB_m(\Gamma)$.

For the purpose of building an encoder, we construct a labeled graph
$G$ that presents $N_m(\Gamma)$. The vertex set is defined as
$V=\bigcup_{i=0}^{m-1}\Sigma^i$. The edges, with labels from $\Sigma$,
are given by
\[ a_0a_1\dots a_{i} \xrightarrow{a_{i+1}} a_0a_1\dots a_{i}a_{i+1},\]
for all $0\leq i\leq m-2$ and $a_j\in\Sigma$ for all $j$, as well as
\[ a_0 a_1\dots a_{m-2} \xrightarrow{a_{m-1}} a_1\dots a_{m-2}a_{m-1},\]
for all $a_0a_1 \dots a_{m-2}a_{m-1}\in \cB(\Gamma)$ and
$a_j\in\Sigma$ for all $j$.

It is easily observed that every path of length $m-1$ labeled by
$\omega\in\Sigma^{m-1}$ ends in the vertex labeled by $\omega$. From
then on, by simple induction, assuming $\omega'\omega$ is a label of a
path with $\omega\in\Sigma^{m-1}$, then the path ends in the vertex
$\omega$ and a letter $a\in\Sigma$ may be generated following that
path if and only if $\omega a\in \cB(\Gamma)$.

\begin{theorem}
  \label{th:sw1}
  Let $\Gamma$ be a convex fat SCS. Then for any $\Gamma$-admissible
  $\epsilon>0$, and for all $m\geq k$,
  \[ N_m(\Gamma_{\epsilon}) \esubseteq \cB(\Gamma).\]
\end{theorem}

\begin{IEEEproof}
  Consider $\omega\in N_m(\Gamma_{\epsilon})$, $\omega=a_1 a_2 \dots
  a_n$, $a_i\in\Sigma$, and assume $\abs{\omega}=n \geq 3m-2$. We define
  the $i$th length-$m$ window sliding over $\omega$ as
  \[ \omega_i = a_i a_{i+1}\dots a_{i+m-1},\]
  for all $1\leq i\leq n-m+1$. We conveniently denote
  \[ \mu_i = \fr_{\omega_i}^k \in \Gamma_{\epsilon}.\]
  We also define
  \[ \mu = \frac{1}{n-m+1} \sum_{i=1}^{n-m+1} \mu_i.\]
  Since $\Gamma$ is convex, so is $\Gamma_{\epsilon}$, and therefore
  $\mu\in\Gamma_{\epsilon}$.

  For any $\phi\in\Sigma^k$, the number of occurrences of $\phi$ in
  $\omega_i$ is exactly $(m-k+1)\mu_i(\phi)$. By taking the sum
  $(m-k+1)\sum_{i=1}^{n-m+1}\mu_i(\phi)$ we are overcounting the
  number of times $\phi$ occurs in $\omega$.  However, we note that
  any occurrence of $\phi$ that is fully contained within $a_{m}
  a_{m+1}\dots a_{n-m+1}$ (i.e., within the windows $\omega_m,
  \omega_{m+1},\dots,\omega_{n-2m+2}$), is overcounted by a factor of
  $m-k+1$ since it appears within exactly $m-k+1$ consecutive
  length-$m$ windows $\omega_i$. It follows that
  \begin{align*}
    \fr_{\omega}^k(\phi) &\leq \frac{1}{n-k+1}
    \sum_{i=m}^{n-2m+2} \mu_i(\phi) \\
    &\quad\ +\frac{m-k+1}{n-k+1}\parenv{\sum_{i=1}^{m-1}\mu_i(\phi)+\sum_{n-2m+3}^{n-m+1}\mu_i(\phi)} \\
    &= \frac{n-m+1}{n-k+1}\mu(\phi) \\
    &\quad\ +\frac{m-k}{n-k+1}\parenv{\sum_{i=1}^{m-1}\mu_i(\phi)+\sum_{n-2m+3}^{n-m+1}\mu_i(\phi)} \\
    &\leq \mu(\phi) + \frac{(m-k)(2m-2)}{n-k+1}.
  \end{align*}
  
  On the other hand, a lower bound may be obtained by assuming a maximal
  overcounting factor of $m-k+1$ for all occurrences of $\phi$, regardless
  of position within $\omega$. This time,
  \begin{align*}
    \fr_{\omega}^k(\phi) &\geq \frac{1}{n-k+1} \sum_{i=1}^{n-m+1}\mu_i(\phi)\\
    &= \frac{n-m+1}{n-k+1}\mu(\phi) \\
    &\geq \mu(\phi)-\frac{m-k}{n-k+1}.
  \end{align*}

  We observe that
  \[ 0\leq \frac{m-k}{n-k+1} \leq \frac{(2m-2)(m-k)}{n-k+1}.\]
  Thus, if we define
  \begin{equation}
    \label{eq:skmn}
    S(k,m,n) = \frac{(2m-2)(m-k)}{n-k+1},
  \end{equation}
  then
  \[
  \abs{\fr_{\omega}^k(\phi)-\mu(\phi)}\leq S(k,m,n).
  \]
  
  Let us now define
  \[ N_\epsilon = \frac{(2m-2)(m-k)}{\epsilon}+k-1.\]
  Then for all $n> \max\mathset{N_\epsilon,3m-2}$ and all $\omega\in
  N_m(\Gamma_{\epsilon})$, $\abs{\omega}=n$, we also have
  $\fr_\omega^k\in\Gamma$, i.e., $\omega\in \cB(\Gamma)$. Hence,
  $N_m(\Gamma_{\epsilon}) \esubseteq \cB(\Gamma)$ as claimed.
\end{IEEEproof}

Note that unlike Construction A, here we obtain that
$N_m(\Gamma_{\epsilon})\subseteq^e \cB(\Gamma)$ for every $m\geq
k$. We also note $N_{\epsilon}=\Omega(\frac{1}{\epsilon})$.

A stronger statement than that of Theorem \ref{th:sw1} may be made in
the case WSCS, in which $\epsilon$ is removed. This is due to the fact
that the quantity $S(k,m,n)$ defined  by equation \eqref{eq:skmn} in the proof of Theorem \ref{th:sw1} is in fact
$o(1)$ for constant $k$ and $m$.

\begin{corollary}
  \label{cor:sw2}
  Let $\Gamma$ be a convex fat SCS, fix $m\geq k$, and define the tolerance
  function $\xi(n) = S(k,m,n)$, where $S(k,m,n)$ is defined in
  \eqref{eq:skmn}. Then
  \[ N_m(\Gamma) \esubseteq \overline{\cB}(\Gamma).\]
\end{corollary}
\begin{IEEEproof}
  The proof follows from the proof of Theorem \ref{th:sw1}, by noting
  that $S(k,m,n)$ is both $o(1)$ and $\Omega(\frac{1}{n})$.
\end{IEEEproof}

\begin{theorem}
  \label{th:belowwin}
  Let $\Gamma$ be a closed convex fat SCS. Then
  \[\limsup_{m\to\infty} \ccap(N_m(\Gamma))=\ccap(\cB(\Gamma)).\]
\end{theorem}

\begin{IEEEproof}
  By Corollary \ref{cor:sw2} and Theorem \ref{th:prevpaper}
  \[\ccap(N_m(\Gamma)))\leq\ccap(\overline{\cB}(\Gamma))=\ccap(\cB(\Gamma)).\]
  Note that this statement does not require taking $m$ to infinity,
  and it applies to all $m\geq k$.
	
  For the other direction, we contend that for every
  $\Gamma$-admissible $\epsilon>0$ there exists $M_\epsilon$ such
  that for all $m> M_\epsilon$
  \[
  \ccap\parenv{N_{m^2}\parenv{\Gamma}} \geq \ccap(R_m(\Gamma_{\epsilon})).
  \]
  To prove this claim, let $\omega=\omega_1\omega_2\dots \omega_{\ell}
  \in R_m(\Gamma_{\epsilon})$ with $\ell\geq m$ and
  $\omega_i\in\cB_m(\Gamma_{\epsilon})$ for all $i$. Denote
  $\mu_i=\fr_{\omega_i}^k\in\Gamma_{\epsilon}$. Let $\omega'$ be any
  length-$m^2$ subword of $\omega$, and let us check the frequency any
  $k$-tuple $\phi\in\Sigma^{k}$ appears in it.
	
  Such a sequence $\omega'$ is surely fully contained in some $m+1$
  consecutive subwords, say $\omega_j \omega_{j+1}\dots \omega_{j+m}$. Let
  us denote
  \[ \mu=\frac{1}{m+1}\sum_{i=0}^{m} \mu_{j+i}.\]
  Again, $\mu\in \Gamma_\epsilon$ due to the convexity of
  $\Gamma_\epsilon$.
  
  In a similar fashion to previous proofs, the frequency of $\phi$ in
  $\omega'$ is easily seen to be upper bounded by
  \begin{align*}
    \fr_{\omega'}^k(\phi) & \leq
    \frac{(m-k+1)\sum_{i=0}^m \mu_{j+i}(\phi)+(k-1)m}{m^2-k+1} \\
    & \leq \mu(\phi)+ \frac{m}{m^2-k+1},
  \end{align*}
  by accounting for the occurrences of $\phi$ in each subword
  $\omega_{j+i}$, and upper bounding the effect of the $m$
  concatenation points between those subwords.

  Conversely, the $m-1$ subwords
  $\omega_{j+1}\omega_{j+2}\dots\omega_{j+m-1}$ must be fully
  contained within $\omega'$. Thus, we obtain the lower bound
  \begin{align*}
  	\fr_{\omega'}^k(\phi)&\geq \frac{(m-k+1)\sum_{i=1}^{m-1}\mu_i(\phi)}{m^2-k+1} \\
  	& \geq  \mu(\phi)-\frac{mk+2(m-k+1)}{m^2-k+1}.
  \end{align*}
  It therefore follows that
  \begin{equation}
    \label{eq:m2}
    \abs{\fr_{\omega'}^k(\phi)-\mu(\phi)} \leq \frac{mk+2(m-k+1)}{m^2-k+1}.
  \end{equation}

  Since the right-hand side of \eqref{eq:m2} is $o(1)$ as
  $m\to\infty$, there exists $M_\epsilon\geq k$ such that for all
  $m>M_\epsilon$, $\fr_{\omega'}^k\in\Gamma$, and thus
  \[R_m(\Gamma_{\epsilon})\subseteq
  N_{m^2}\parenv{\Gamma}.\]
  as claimed. It follows that for $m> M_\epsilon$,
  \[
  \ccap\parenv{N_{m^2}\parenv{\Gamma}} \geq \ccap(R_m(\Gamma_{\epsilon})).
  \]
  Taking $\limsup_{m\to\infty}$ on both sides we obtain
  \begin{align*}
    \limsup_{m\to\infty}\ccap\parenv{N_{m}\parenv{\Gamma}} &\geq
    \limsup_{m\to\infty}\ccap\parenv{N_{m^2}\parenv{\Gamma}}\\
    &\geq \limsup_{m\to\infty}\ccap(R_m(\Gamma_{\epsilon}))\\
    & = \ccap(\cB(\Gamma_{\epsilon})),
  \end{align*}
  where the last equality is due to Theorem \ref{th:rep2}. Now, since
  this holds for all $\Gamma$-admissible $\epsilon>0$, taking the
  limit as $\epsilon\to 0$, by the continuity guaranteed in Theorem
  \ref{th:prevpaper} we get  \[\limsup_{m\to\infty}\ccap\parenv{N_{m}\parenv{\Gamma}} \geq \ccap(\cB(\Gamma)),\]
  which completes the proof.
\end{IEEEproof}

The graph $G$ that presents $N_m(\Gamma)$, as described above, is
$(m,0)$-definite, i.e., all the paths that generate a given word of
length $m+1$ symbols agree on the edge that generated the last
symbol. The graph is not necessarily an encoder (due to an unequal
out-degree), but by using the state-splitting algorithm on $G$ we may
generate a $p:q$ encoder.

Apart from describing two constructions for encoders, we have thus
far also proved the following corollary.

\begin{corollary}
  Let $\Gamma$ be a closed convex fat SCS. Then
  \[ \ccap^\subseteq(\Gamma) = \ccap(\cB(\Gamma)).\]
\end{corollary}
\begin{IEEEproof}
  This is immediate either from Corollary \ref{cor:belowblock} or from Theorem
  \ref{th:belowwin}.
\end{IEEEproof}

\subsection{A Short Case Study}

As a short case study we provide the following example. Consider the
SCS over $\Sigma=\mathset{0,1}$, which is defined by the set
\[\Gamma=\mathset{\mu\in\cP(\Sigma^2) ~:~ \mu(11)\leq 0.205}.\] 
This SCS was called the $(0,1,0.205)$-RLL SCS in \cite{EliMeySch16},
and its capacity is $\ccap(\cB(\Gamma))\approx 0.98$. We investigate
the encoders presented thus far, with an intention of building an
encoder with rate $\frac{3}{4}$.

We first focus on the block encoder associated with
$R_m(\Gamma)$. Choosing $\epsilon=0.005$, a quick use of
\eqref{eq:MeA} shows that any $m>200$ guarantees that we satisfy the
semiconstraints. A finer analysis, accounting for divisibility
conditions, reveals all $m>156$ suffice. The latter is indeed tight,
since for $m=156$ we have $\omega=1^{32}0^{123}1\in
\cB_{156}(\Gamma_{0.005})$, but
\[\lim_{i\to\infty} \fr_{\omega^i}^2(11)=\frac{32}{156}>0.205,\]
so for large enough $i$, $\omega^i$ does not satisfy the
semiconstraints. However, there exist smaller values of $m$ which are
acceptable. The smallest one is $m=5$. However, in this case,
$\abs{\cB_5(\Gamma_{0.005})}=13$, not achieving the required rate of
$\frac{3}{4}$. The next possible acceptable value is $m=10$, in which
case $\abs{\cB_{10}(\Gamma_{0.005})}=379$, exceeding the required rate,
but at the cost of having an unwieldy number of edges in the encoder.

On the other hand, the encoder associated with $N_m(\Gamma)$ is
simpler. We can choose $m=6$. We first construct the modified
De-Bruijn graph of order $m-1=5$ where we allow at most a single
appearance of the pattern $11$. Since we would like to build an
encoder with rate $\frac{3}{4}$, we take the graph to its $4$th power,
and keep the appropriate irreducible subgraph. After combining
vertices with the same follower sets and applying the state-splitting
algorithm, we obtain an encoder with $14$ vertices and $112$ edges.

\section{Approaching Capacity from Above}
\label{sec:above}

In this section we consider the dual question to the one asked in
Section \ref{sec:below}: which fully constrained systems, presented as
the language of a directed labeled graph, \emph{contain} a given
semiconstrained system. Additionally, we would like to know whether
the capacity of a sequence of those fully constrained system can
approach the capacity of the semiconstrained system in the limit.
In an analogous fashion to Definition \ref{def:below}, we define
the following.

\begin{definition}
  \label{def:above}
  Let $\Gamma\subseteq\cP(\Sigma^k)$ be a SCS.  The \emph{fully
    constrained containing capacity} is defined as

 \begin{equation}
\label{eq:ccap_supset}
 \ccap^\supseteq(\Gamma) = \inf_{\cL(G)\supseteq\cB(\Gamma)} \ccap(\cL(G)).
\end{equation}
\end{definition}

As we shall soon see, the result is quite pessimistic. We first give
an auxiliary lemma, and then proceed to prove the main theorem. For this
lemma we require the following definition.

\begin{definition}
  \label{def:graph}
  Let $\eta\in\cP(\Sigma^k)$ be a rational measure. We define the
  following graphs $nG_\eta$, for each $n\in\N$. Let $M\in\N$ be the
  smallest natural number such that $M\eta$ is an integer vector. The
  initial vertex set of $nG_\eta$ is $V=\Sigma^{k-1}$. For each
  $a_0,a_1,\dots,a_{k-1}\in\Sigma$, we place $nM\eta(a_0 a_1 \dots
  a_{k-1})$ parallel edges from vertex $a_0 a_1 \dots a_{k-2}$ to
  vertex $a_1 a_2 \dots a_{k-1}$. Finally, we remove vertices with
  zero in-degree and out-degree, i.e., isolated vertices.
\end{definition}

\begin{lemma}
  \label{lem:prefix}
  Let $\eta\in\cPsi$ be a positive (entry-wise) rational and
  shift-invariant measure. Then for any $\alpha\in\Sigma^*$ there
  exists $\beta\in\Sigma^*$ such that $\fr_{\alpha\beta}^k=\eta$.
\end{lemma}

\begin{IEEEproof}
  Since $\eta$ is rational and shift invariant, for all
  $a_0,a_1,\dots,a_{k-1}\in\Sigma$ we have that $\eta(a_0 a_1\dots
  a_{k-1})\in\Q$, and
  \[ \sum_{b\in \Sigma}\eta(a_0 a_1\dots a_{k-2} b) = \sum_{b\in \Sigma}\eta(b a_0 a_1\dots a_{k-2}).\]
  Assume we are given a sequence $\alpha\in\Sigma^m$ with $m\geq k$
  (if $m<k$ we arbitrarily extend $\alpha$ so its length is at least
  $k$).

  We now consider the graph $(m+1)G_\eta$. We note that since $\eta$
  is positive, the graph $(m+1)G_\eta$ is strongly connected, i.e.,
  there is a directed path between any source vertex and destination
  vertex.  Additionally, the shift-invariance property of $\eta$
  implies that the in-degree of every vertex equals its out-degree.
	
  With a directed path of $n$ edges in the graph we associate a
  sequence of length $n+k-1$ over $\Sigma$ via a sliding-window
  reading of the sequence. Formally, a sequence $\alpha=a_0 a_1 \dots
  a_{n+k-2}\in\Sigma^{n+k-1}$, $a_i\in\Sigma$, is associated with the
  directed path whose $i$th edge is $a_i a_{i+1}\dots a_{i+k-2} \to
  a_{i+1} a_{i+2}\dots a_{i+k-1}$, for all $0\leq i\leq n-1$. Since
  this mapping is a bijection (up to parallel edges), by abuse of
  notation we shall refer to $\alpha$ as both the sequence and the
  path.
	
  The given sequence $\alpha$ describes a path in $(m+1)G_\eta$, where
  the graph parameter $(m+1)$ ensures the path can consist of distinct
  (though perhaps parallel) edges. Let us remove the edges of this
  path from the graph to obtain a graph $G'$.
	
  First we note that $G'$ is still strongly connected since any two
  vertices originally connected by an edge (perhaps several parallel
  edges) are still connected by at least one edge. This is because a
  total of $m-k+1<m+1$ edges were removed, and two vertices connected
  by an edge in $(m+1)G_\eta$ are actually connected by at least $m+1$
  edges.
	
  Next we distinguish between two cases. If the removed path $\alpha$
  was a cycle, then $G'$ still has the property that every vertex has
  an equal in-degree and out-degree. Thus, it contains an Eulerian
  cycle (going over every edge exactly once) which we denote as
  $\beta$. Since the ending vertex of $\alpha$ still has a positive
  out-degree, we may, without loss of generality, choose $\beta$ to
  begin in the same vertex. It follows that $\alpha\beta$ is an
  Eulerian cycle in the original graph $(m+1)G_\eta$.
	
  In the second case, the removed path $\alpha$ is not a cycle. In
  that case, except for the starting vertex and ending vertex of
  $\alpha$, all other vertices have equal in-degree and
  out-degree. The starting vertex of $\alpha$ has an in-degree larger
  by $1$ compared with its out-degree, and the ending vertex of
  $\alpha$ has the reversed situation. Thus, $G'$ contains an Eulerian
  path $\beta$ that starts in the ending vertex of $\alpha$, and ends
  in the starting vertex of $\alpha$. Again, $\alpha\beta$ is
  therefore an Eulerian cycle in the original graph $(m+1)G_\eta$.
	
  We note that the sequence associated with the sliding-window reading
  induced by the path $\alpha\beta$ in $(m+1)G_\eta$ has each window
  $\phi\in\Sigma^k$ appear exactly as an $\eta(\phi)$ fraction of the
  windows of size $k$, i.e.,
  $\fr_{\alpha\beta}^k(\phi)=\eta(\phi)$.
\end{IEEEproof}

\begin{corollary}
  \label{cor:prefix}
  Let $\Gamma$ be a fat SCS.  Then for all $\alpha\in\Sigma^*$ there
  exists $\beta\in\Sigma^*$ such that $\alpha\beta\in \cB(\Gamma)$,
  i.e., any finite prefix may be completed to a word in the
  semiconstrained system.
\end{corollary}

\begin{IEEEproof}
  Since $\Gamma$ is fat, there exists a probability measure
  $\nu\in\sint(\Gamma\cap\cPsi)$, i.e., $\nu$ is shift invariant and
  in the interior of $\Gamma$. We can take a sequence of rational
  shift-invariant probability measures that converge to $\nu$, and
  since there exists an $\epsilon>0$ environment of $\nu$ contained
  within $\Gamma\cap\cPsi$, then we deduce the existence of a rational
  shift-invariant probability measure $\eta\in\Gamma$. From Lemma
  \ref{lem:prefix} we can find $\beta\in\Sigma^*$ such that
  $\alpha\beta\in\cB(\Gamma)$.
\end{IEEEproof}

We note that the property that every prefix may be extended to a word
in $\cB(\Gamma)$, is called \emph{right density} in formal-language
theory (e.g., see \cite{Shy86}).

We now state the main result of this section.

\begin{theorem}
  \label{th:fullcon}
  Let $\Gamma$ be a fat SCS. Let $\Gamma'$ be a fully constrained
  system such that $\cB(\Gamma)\subseteq \cB(\Gamma')$. Then
  $\cB(\Gamma')=\Sigma^*$, and thus
  \[ \ccap^\supseteq(\Gamma) = \log_2 \abs{\Sigma}.\]
\end{theorem}

\begin{IEEEproof}
  Let $G$ be a directed labeled graph that presents $\cB(\Gamma')$,
  i.e., $\cL(G)=\cB(\Gamma')$ and the edges are labeled by
  $\Sigma$. By Corollary \ref{cor:prefix}, for any
  $\alpha\in\Sigma^*$, there exists $\beta\in\Sigma^*$ such that
  $\alpha\beta\in \cB(\Gamma)$, and therefore also $\alpha\beta\in
  \cB(\Gamma')$. Thus, the graph $G$ has a directed path whose labels
  generate $\alpha\beta$. It follows that a prefix of the path that
  generates $\alpha\beta$, generates $\alpha$. Hence,
  $\cL(G)=\Sigma^*$.
\end{IEEEproof}

We note the peculiar asymmetry between fully constrained systems
\emph{contained} within a fat SCS, and fully constrained systems
\emph{containing} a fat SCS. While in the former we have a sequence of
such fully constrained systems that approach the capacity of the given
fat SCS, in the latter there is exactly one fully constrained system
containing the SCS, and that is the entire space $\Sigma^*$.

\section{Combining SCS with combinatorial constraints}
\label{sec:RF}

In our discussion thus far, we required any SCS $\Gamma$ to be
fat. Unfortunately, if for some $\phi\in\Sigma^k$ we have
$\mu(\phi)=0$ for all $\mu\in\Gamma$, then $\Gamma$ is not
fat. Intuitively, in that case $\Gamma$ does not occupy all the
dimensions of $\Sigma^k$. It follows that none of the results obtained
in the previous sections apply to fully constrained systems, since the
latter employ such zero constraints, and are therefore not fat SCS.
In the following we discuss how this situation can be resolved by
defining relatively-fat SCS.

We start by defining the set of forbidden words, that is, those words
which are never subwords of the admissible words of the given SCS.

\begin{definition}
Let $\Gamma\subseteq \cP(\Sigma^k)$ be a set of probability
measures. We denote by $\cF(\Gamma)\subseteq\Sigma^k$ the following
set of \emph{$\Gamma$-forbidden} $k$-tuples,
\[\cF(\Gamma)=\mathset{\phi\in\Sigma^k ~:~ \forall \mu\in\Gamma, \mu(\phi)=0}.\]
\end{definition}

Informally, we call a set $\Gamma$ \emph{relatively fat (RF)} if apart
from $\cF(\Gamma)$ it is fat. A formal definition follows.

\begin{definition}
  For any $\Gamma\subseteq \cP(\Sigma^k)$, let us denote
  $\cD=\Sigma^k\setminus \cF(\Gamma)$. We say that
  $\Gamma$ is \emph{relatively fat (RF)} if
  \[\scl_{\cD}(\sint_{\cD}(\Gamma\cap\cP_{\mathrm{si}}(\cD)))=\scl_{\cD}(\Gamma \cap \cP_{\mathrm{si}}(\cD)),\]
  where $\scl_{\cD}$ and $\sint_{\cD}$ are the closure and interior
  with respect to $\cD$, respectively, and $\cP_{\mathrm{si}}(\cD)$
  denotes the set of shift-invariant measures on $\cD$.
\end{definition}

We mention briefly that other definitions of RF sets are possible,
somewhat generalizing the definition we use. For example, one may
generalize the definition to one that requires $\Gamma$ to be
contained within an affine space of dimension possibly lower than
$\cP(\Sigma^k)$, and further that $\Gamma$ is fat with respect to that
affine space. However, such generality is not required by us.

We now go through the results obtained thus far for fat SCS, and
describe the necessary changes required to make them work for RF SCS
as well. We first note that the general form of Theorem
\ref{th:prevpaper}, which holds for every set $\Gamma$ and is given in \cite[Ch. 3]{DemZei98} is as follows (with modified notations).
\begin{theorem}
  \label{th:capcomp}
  Let $\Gamma\subseteq\cP(\Sigma^k)$ be closed and convex. Then
  \begin{multline*}
    \log_2 \abs{\Sigma} - \inf_{\eta\in\sint(\Gamma\cap\cPsi)}H(\eta|\eta')\leq  \ccap(\cB(\Gamma))
    \\ \leq\log_2 \abs{\Sigma} - \inf_{\eta\in\scl(\Gamma\cap\cPsi)}H(\eta|\eta'),
  \end{multline*}
  where $\sint$ and $\scl$ are the interior and closure of a set.
\end{theorem}

In the case of RF SCS, the interior of $\Gamma$ may be empty and
therefore Theorem \ref{th:capcomp} states that the capacity of a RF
SCS is bounded from below by $-\infty$. We are therefore
left with the upper bound of
\[\ccap(\cB(\Gamma))\leq\log_2 \abs{\Sigma} - \inf_{\eta\in\scl(\Gamma\cap\cPsi)}H(\eta|\eta').\] 
%
%


Construction \ref{con:A} does not work for RF SCS. The cause of
failure is the obvious concatenation point between blocks, which may
contain a forbidden word. For example, assume a binary alphabet,
$k=2$, and $\cF=\mathset{11}$, i.e., the SCS is in fact the
$(1,\infty)$-RLL fully-constrained system. In this case, taking two
words of length $m$, the first, $\alpha_1$, ending with a $1$, and the
second $\alpha_2$, starting with a $1$, and concatenating them
together, will create the forbidden pattern $11$ in $\alpha_1
\alpha_2$. A way of solving this problem is by placing a carefully
crafted string, $\beta$, between the two blocks, i.e., $\alpha_1 \beta
\alpha_2$. As long $\abs{\beta}=o(m)$, Construction \ref{con:A} works.
A similar method, developed for fully-constrained multidimensional RLL
systems was described in \cite{Etz97}.

In comparison, Construction \ref{con:B} indeed does work for RF
SCS. The only change needed in the proofs is to alter the definition
of $\Gamma_{\epsilon}$ from \eqref{eq:gammaeps} to
\[\Gamma_{\epsilon}=\mathset{\eta\in\Gamma : \forall \phi\in\Sigma^k\setminus\cF(\Gamma), \inf_{\mu\in\Gamma^c} \abs{\eta(\phi)-\mu(\phi)}>\epsilon}.\]


Having considered the generalization of Section \ref{sec:below} to RF
SCS, we now turn to discuss the generalization of Section
\ref{sec:above}. In what follows, we do not even require the
relative-fatness property. We are therefore interested in fully
constrained systems containing a given SCS. Unlike contained fully
constrained system, in the containing case the discussion is somewhat
more involved.

We recall some useful notions from graph theory. Two vertices, $v_1$
and $v_2$, in a directed graph $G$, are said to be \emph{bi-connected}
if there is a directed path from $v_1$ to $v_2$, and a directed path
from $v_2$ to $v_1$.  Bi-connectedness is an equivalence relation, and
its equivalence classes are called \emph{strongly connected components}.

In Corollary \ref{cor:prefix} we used the fact that for a fat
$\Gamma$, there exists a rational $\eta\in\Gamma$ such that $G_\eta$
is a single strongly connected component. Unfortunately, this is no
longer the case for general SCS, even if we restrict ourselves to RF
SCS as shown in the following example.

\begin{example}
  \label{ex:nonconv}
  Fix $\Sigma=\mathset{0,1}$, and $k=4$. Define
  $\mu_1,\mu_2\in\cPsi$ as follows:
  \[\mu_1=\delta_{1111}, \qquad\text{ and}\qquad
  \mu_2=\frac{1}{2}(\delta_{1010}+\delta_{0101}),\] where
  $\delta_\phi$ denotes probability measure of value $1$ at $\phi$,
  and $0$ elsewhere. Let $\Gamma$ be the convex hull of $\mu_1$ and
  $\mu_2$. Then $\Gamma$ is non-empty, convex, relatively fat, and
  contains only shift-invariant measures.  However, except for $\mu_1$
  and $\mu_2$, there is no other $\eta\in\Gamma$ such that $G_{\eta}$
  has a single strongly connected component.
\end{example}

The following definition is intended to capture and isolate this kind of pathological behaviour.

\begin{definition}
  Let $\Gamma\subseteq\cPsi$ be a SCS. The \emph{essential part} of
  $\Gamma$ is defined as
  \[\ess(\Gamma)=\mathset{\eta\in\Gamma ~:~ \cB(\mathset{\eta})\neq\emptyset }.\]
\end{definition}

Thus, $\ess(\Gamma)$ keeps only those measures of $\Gamma$ that have
at least one admissible word. Note that by definition, $\ess(\Gamma)$
contains only rational measures. We also note that even if $\Gamma$ is
convex, the set $\ess(\Gamma)$ may not necessarily be convex (even if
we consider only convex rational combinations of measures from
$\ess(\Gamma)$). This can be seen in Example \ref{ex:nonconv}, in
which $\ess(\Gamma)=\mathset{\mu_1,\mu_2}$.

\begin{lemma}
  \label{lem:connect}
  Let $\eta\in\cPsi$ be a rational shift-invariant measure. Then
  $\cB(\mathset{\eta})\neq \emptyset$ if and only if $G_\eta$ is
  strongly connected after removing isolated vertices.
\end{lemma}
\begin{IEEEproof}
  In the first direction assume
  $\cB(\mathset{\eta})\neq\emptyset$. Then let
  $\omega\in\cB(\mathset{\eta})$. As already shown in the proof of
  Lemma \ref{lem:prefix}, $\omega$ corresponds to an Eulerian cycle in
  $nG_\eta$ for some $n\in\N$. Thus, $nG_\eta$ is strongly connected,
  and since $n$ does not affect this property, $G_\eta$ is also
  strongly connected.

  In the other direction, assume $G_\eta$ is strongly connected. Since
  $\eta$ is shift invariant, the in-degree and out-degree of each
  vertex are equal, and there exists an Eulerian cycle in
  $G_\eta$. Again, by the proof of Lemma \ref{lem:prefix}, this cycle
  corresponds to a word $\omega\in\Sigma^*$ with
  $\fr^k_\omega=\eta$. Thus, $\omega\in\cB(\mathset{\eta})$.
\end{IEEEproof}

The next step we take is to define the essential graph of a SCS.

\begin{definition}
  Let $\Gamma\subseteq\cPsi$. Denote by $G_{\ess}(\Gamma)$ the
  following directed labeled graph: Vertices are represented by
  elements of $\Sigma^{k-1}$. For each $\phi=a_0 a_1 \dots
  a_{k-1}\in\Sigma^k$, such that there exists some
  $\eta\in\ess(\Gamma)$ with $\eta(\phi)>0$, we place an edge $a_0
  a_1\dots a_{k-2} \to a_1 a_2 \dots a_{k-1}$, labeled by $a_{0}$.
  Any isolated vertices (i.e., vertices with both in-degree and
  out-degree of zero) are then removed.
\end{definition}

Intuitively, $G_{\ess}(\Gamma)$ is the union of all $G_\eta$,
$\eta\in\ess(\Gamma)$, where parallel edges are merged, and isolated
vertices are removed. Here, we define the union of two directed
labeled graphs, $G_1=(V_1,E_1)$, and $G_2=(V_2,E_2)$, with
$E_i\subseteq V_i\times V_i\times \Sigma$, as $G_1\cup G_2=(V_1\cup
V_2, E_1\cup E_2)$. We note that the sets of vertices of the two
original graphs are not necessarily disjoint, and the same goes for
the sets of edges. We collect some more insight into words having a
shift-invariant measure.

\begin{lemma}
  \label{lem:cycle}
  Let $\eta\in\cPsi$ be a shift-invariant measure, and let $\omega=a_0
  a_1 \dots a_{n-1}\in\Sigma^n$ be a word for which
  $\fr_{\omega}^k=\eta$. Then $a_0\dots a_{k-2}=a_{n-k+1}\dots
  a_{n-1}$, i.e., the $(k-1)$-prefix and $(k-1)$-suffix of $\omega$
  are equal.
\end{lemma}

\begin{IEEEproof}
  Let $\omega$ be a word with $k$-tuple distribution given by
  $\eta$. Since $\eta$ is shift invariant, for every
  $\phi\in\Sigma^{k-1}$ we have
  \begin{equation}
    \label{eq:si}
    \sum_{a\in\Sigma}\eta(a\phi)=\sum_{a\in\Sigma}\eta(\phi a).
  \end{equation}
  In particular, let us examine $\phi=a_0 a_1 \dots a_{k-2}$, the
  $(k-1)$-prefix of $\omega$. The left-hand side of \eqref{eq:si} is
  given by
  \begin{equation}
    \label{eq:si1}
    \frac{\abs{\mathset{(\alpha,\gamma) ~:~ \alpha,\gamma\in\Sigma^*, \alpha\phi\gamma=\omega}}-1}{n-k+2},
  \end{equation}
  where the subtraction of $1$ in the numerator is due to the fact we
  require a letter to appear before $\phi$, and therefore cannot count
  its appearance as a prefix of $\omega$.

  Assume to the contrary that $\phi$ is not the $(k-1)$-suffix of $\omega$.
  In that case, every occurrence of $\phi$ in $\omega$ is followed by
  a letter, and then the right-hand side of \eqref{eq:si} is
  \[ \frac{\abs{\mathset{(\alpha,\gamma) ~:~ \alpha,\gamma\in\Sigma^*, \alpha\phi\gamma=\omega}}}{n-k+2},\]
  but that differs from \eqref{eq:si1}, a contradiction.
\end{IEEEproof}

We shall call a word $\omega\in\Sigma^*$ \emph{$k$-shift-invariant}
if $\fr^k_\omega\in\cPsi$. By the previous lemma, the $(k-1)$-suffix
of $\omega$ equals its $(k-1)$-prefix. We shall therefore find it convenient
to chop off the $(k-1)$-suffix of $\omega$ using the following operator.
If $\omega=a_0 a_1 \dots a_{n-1}$, $n\geq k-1$, then we define
\[ \suffchop_{k-1}(\omega)= a_0 a_1 \dots a_{n-k}.\]
The following corollary is therefore immediate.

\begin{lemma}
  \label{lem:iness}
  Let $\Gamma\subseteq\cPsi$ be a SCS. Then for every
  $\omega\in\cB(\Gamma)$ there exists a cycle in $G_{\ess}(\Gamma)$
  generating $\suffchop_{k-1}(\omega)$.  Hence,
  \[ \cB(\Gamma)\subseteq \cL(G_{\ess}(\Gamma)).\]
\end{lemma}
\begin{IEEEproof}
  Let $\omega\in\cB(\Gamma)$, and denote $\abs{\omega}=n$. We can assume $n\geq k-1$. Additionally, we must have $\eta=\fr^k_{\omega}\in
  \ess(\Gamma)$, by definition. If we read $\omega$ by a sliding
  window of size $k-1$, then by Lemma \ref{lem:cycle} we get a
  sequence of vertices of $G_{\ess}(\Gamma)$ forming a cycle. The labels
  along this cycle generate $\suffchop_{k-1}(\omega)$. We can then
  take again the first $k-1$ edges of the cycle to complete a reading
  of $\omega$. Thus, $\omega\in\cL(G_{\ess}(\Gamma))$.
\end{IEEEproof}

We argue that every word obtained by reading the labels of edges along
a walk on $G_{\ess}(\Gamma)$ can be completed to a word in
$\cB(\Gamma)$.

\begin{theorem}
  \label{th:comp2}
  Let $\Gamma\in\cPsi$ be a convex SCS. Then, for every
  $\alpha\in\cL(G_{\ess}(\Gamma))$ there exists $\beta\in\Sigma^*$
  such that $\alpha\beta\in\cB(\Gamma)$.
\end{theorem}

\begin{IEEEproof}
  Let $\alpha\in\cL(G_{\ess}(\Gamma))$, $\abs{\alpha}=n$, be a word
  that is obtained by reading the labels of edges $e_1\to e_2 \to
  \dots \to e_{n}$ along a path in $G_{\ess}(\Gamma)$.

  Each edge $e_i$ corresponds to some $\phi_i\in\Sigma^k$.  By the
  definition of $G_{\ess}(\Gamma)$, the edge $e_i$ exists since there
  exists $\eta_i\in\Gamma$, such that $\eta_i(\phi_i)>0$ and
  $\cB(\eta_i)\neq\emptyset$. Thus, $\eta_i$ is rational and shift
  invariant. By Lemma \ref{lem:connect}, $G_{\eta_i}$ is strongly
  connected (after removing isolated vertices\footnote{Throughout the
    proof we remove isolated vertices from graphs.}).

  We now take a convex combination
  \[ \eta = \sum c_i \eta_i,\]
  where $c_i>0$, $c_i\in\Q$, for all $1\leq i\leq n$, and
  $\sum_{i=1}^n c_i=1$. Since $\Gamma$ is convex, we have
  $\eta\in\Gamma$. By our previous observations, $G_\eta$ contains the
  path $e_1\to\dots e_n$, and is the union of the graphs
  $\mathset{G_{\eta_i}}_{i=1}^n$, each of which is strongly
  connected. Thus, $G_{\eta_i}$ is also strongly connected.

  We first note that by Lemma \ref{lem:connect},
  $\cB(\mathset{\eta})\neq\emptyset$, i.e., $\eta\in\ess(\Gamma)$.
  Following the same reasoning as in the proof of Lemma
  \ref{lem:prefix}, there exists $m\in\N$ such that there exists an
  Eulerian cycle in $mG_{\eta}$ starting with the path $e_1\to\dots
  e_n$. This Eulerian cycle therefore corresponds to a reading of a
  word $\alpha\beta$, by sliding windows of size $k$, for which
  $\fr^k_{\alpha\beta}=\eta\in\Gamma$. Hence, $\alpha\beta\in\cB(\Gamma)$.
\end{IEEEproof}


As a corollary we obtain the main result of this section.

\begin{corollary}
  Let $\Gamma\subseteq\cPsi$ be a convex SCS. Then $\cL(G_{\ess}(\Gamma))$
  is the unique smallest fully constrained system containing $\cB(\Gamma)$.
  In particular,
  \[\ccap^\supseteq(\Gamma) = \ccap(\cL(G_{\ess}(\Gamma))).\]
\end{corollary}

\begin{IEEEproof}
  Let $G$ be a directed graph, with edges labels from $\Sigma$, such
  that $\cB(\Gamma)\subseteq\cL(G)$. Consider a word
  $\alpha\in\cL(G_{\ess}(\Gamma))$. By Theorem \ref{th:comp2} there
  exists $\beta\in\Sigma^*$ such that
  $\alpha\beta\in\cB(\Gamma)$. Thus, $\alpha\beta\in\cL(G)$.  In
  particular, a prefix of a path generating $\alpha\beta$ in $G$,
  generates $\alpha$. Hence, $\alpha\in\cL(G)$, and
  $\cL(G_{\ess}(\Gamma))\subseteq\cL(G)$. The claims now follow.
\end{IEEEproof}

We devote the remainder of the section for some curious observations.
Our first observations, is that while one might initially assume the
fully constrained containing capacity to be monotone increasing in the
capacity, this is not generally the case, as the following example
shows.

\begin{example}
  \label{ex:mono}
  Let $\Gamma_1,\Gamma_2$ be SCS over $\Sigma=\mathset{0,1}$ with
  $k=3$ defined by
  \begin{align*}
    \Gamma_1&=\mathset{\mu\in\cPsi ~:~ \mu(000),\mu(111),\mu(101)\leq 0.01}. \\ 
    \Gamma_2&=\mathset{\mu\in\cPsi ~:~ \mu(000)=0}.
  \end{align*}

  We note that both $\Gamma_1$ and $\Gamma_2$ are shift invariant,
  convex, and relatively fat. The capacity of $\Gamma_1$ may be
  obtained using Theorem \ref{th:prevpaper}, and that of $\Gamma_2$
  is also easily obtained since it is also a fully constrained system.
  We reach
  \[ \ccap(\cB(\Gamma_1))\approx 0.462 ,\qquad \ccap(\cB(\Gamma_2))\approx 0.879.\]

  One can easily see that $G_{\ess}(\Gamma_1)$ is a complete De-Bruijn
  graph of order $2$. Thus,
  \[ \ccap^\supseteq(\Gamma_1) = 1.\]
  Since $\Gamma_2$ is fully constrained to begin with,
  \[ \ccap^\supseteq(\Gamma_2) = \ccap(\cB(\Gamma_2)) \approx 0.879.\]
  Hence,
  \[\ccap(\cB(\Gamma_1))<\ccap(\cB(\Gamma_2)),\]
  but
  \[\ccap^\supseteq(\Gamma_1)>\ccap^\supseteq(\Gamma_2).\]
\end{example}

While a gap may exist between $\ccap(\cB(\Gamma))$ and
$\ccap^\supseteq(\Gamma)$, as is demonstrated in Example
\ref{ex:mono}, this is never the case with zero capacity.

\begin{lemma}
  \label{lem:zcap}
  Let $\Gamma\subseteq\cPsi$ be a convex SCS. Then we have
  $\ccap(\cB(\Gamma))=0$ if and only if $\ccap^\supseteq(\Gamma)=0$.
\end{lemma}
\begin{IEEEproof}
  If the first direction, assume we have
  $\ccap^\supseteq(\Gamma)=0$. By definition,
  \[\ccap(\cB(\Gamma))\leq \ccap^\supseteq(\Gamma)=0,\]
  and therefore $\ccap(\cB(\Gamma))$ is either $0$ or $-\infty$. We
  note that $\cB(\Gamma)\neq\emptyset$, since then we have
  $\cL(G_{\ess}(\Gamma))=\emptyset$ implying
  $\ccap^\supseteq(\Gamma)=-\infty$, a contradiction. It follows that
  we must have some $\omega\in\cB(\Gamma)$, $\abs{\omega}\geq k$. Denote
  $\omega=\alpha\beta$, where $\alpha=\suffchop_{k-1}(\omega)$. But
  then
  \[\fr^k_{\alpha^n \beta}=\fr^k_{\alpha\beta}\in\Gamma,\]
  for all $n\in\N$, and $\cB(\Gamma)$ is an infinite set, giving us
  the desired $\ccap(\cB(\Gamma))=0$.
  
  In the other direction, assume $\ccap(\cB(\Gamma))=0$. By Lemma
  \ref{lem:iness}, $\cB(\Gamma)\subseteq\cL(G_{\ess}(\Gamma))$. Let
  $\omega\in\cB(\Gamma)$, with $\fr^k_\omega=\eta$. Then $G_\eta$
  (after removing isolated vertices) is Eulerian, since $\eta$ is
  shift invariant. We note that this Eulerian cycle must be simple,
  otherwise $nG_\eta$ contains an exponential (in $n$) number of
  Eulerian cycles, implying $\ccap(\cB(\Gamma))>0$, a contradiction.

  Now, assume $\omega_1,\omega_2\in\cB(\Gamma)$ be two distinct words,
  with $\fr^k_{\omega_i}=\eta_i$.  As in the proof of Theorem
  \ref{th:comp2}, the convexity of $\Gamma$ implies that the simple
  Eulerian cycles of $G_{\eta_1}$ and $G_{\eta_2}$ are either
  identical or disjoint. Otherwise, an appropriate rational convex
  combination of $\eta_1$ and $\eta_2$ results in some $\eta\in\Gamma$
  whose $G_\eta$ is Eulerian and non-simple, implying a positive
  capacity for $\cB(\Gamma)$, a contradiction.

  It follows that $G_{\ess}(\Gamma)$ is a union of disjoint simple Eulerian
  cycles. Thus, $\ccap^\supseteq(\Gamma)=\ccap(\cL(G_{\ess}(\Gamma)))=0$.
\end{IEEEproof}

By the proof of Lemma \ref{lem:zcap} we also observe that for a convex
$\Gamma\subseteq\cPsi$ with $\ccap(\cB(\Gamma))=0$, we have that
$\cB(\Gamma)$ is a regular language, though not necessarily a fully
constrained system. This is no longer true if we omit the requirement
that $\Gamma\subseteq\cPsi$, as the following example shows.

\begin{example}
  Let $\Sigma=\mathset{0,1}$, $k=2$, and define
  \[\Gamma=\mathset{\mu\in \Sigma^2 ~:~ \mu(00)<\frac{1}{2}, \mu(11)<\frac{1}{2}, \mu(01)=0}.\]
  We note that $\Gamma$ is relatively fat, convex, but contains some
  measures which are not shift invariant. Interestingly,
  \[ \cB(\Gamma) = \mathset{0^n 1^n ~:~ n\in\N},\]
  which is not a regular language.
\end{example}

\section{Conclusions and Discussion}
\label{sec:conc}

In an attempt to find connections between SCS and fully constrained
systems, and motivated by the extensive literature on encoders for
fully constrained system, this work was devoted to fully constrained
system either contained or containing a given SCS. Apart from two
encoder constructions, an interesting asymmetry between contained and
containing fully constrained system emerged. Whereas the former
approach the capacity of the given SCS, the latter are generally
bounded away from it.

We suspect cleaner results may be obtained when considering infinite
sequences. This is apparent from the extra care and combinatorics
employed to handle finite words and non-shift-invariant measures. We
leave this study of infinite sequences to a later work.

Another set of open questions raised by this work is the study of
various complexity properties associated with encoders for SCS. In the
two encoders presented here, we briefly mentioned number of states and
edges, as well as anticipation and memory, as important parameters. A
more in-depth study of these and other parameters, as well as
associated bounds and trade-offs between them, will by the subject of
future work.

\bibliographystyle{IEEEtranS}
\bibliography{allbib}

\end{document}